\documentclass[aps,prb,twocolumn,groupedaddress,draft,showpacs,intlimits,amsmath,amssymb,floats]{revtex4}

\usepackage[final]{graphicx}

\begin{document}
 \title{Eliashberg Analysis of Temperature Dependent Pairing Mechanism in d-wave Superconductors: Application to High Temperature Superconductivity.}
 \author{O. Ahmadi, L. Coffey}
 \date{\today}
 \affiliation{Physics Department, Illinois Institute of Technology,
 Chicago, Illinois 60616}
 \begin{abstract}
Results are presented for the temperature and frequency dependence of the 
real and imaginary parts of the diagonal self energy for  a 
d-wave superconductor. An Eliashberg analysis, which has been successful in recent fitting
of superconductor-insulator-superconductor (SIS) tunnel junction conductances 
for Bi$_{2}$Sr$_{2}$CaCu$_{2}$O$_{8}$ (Bi-2212),
is extended to finite temperatures. The superconducting pairing mechanism is assumed to originate in the spin fluctuations of the copper-oxide 
planes, and is modelled by a function incorporating the 
spin resonance mode measured at an 
energy of approximately 40 meV for optimally doped Bi-2212. The effect of the temperature dependence of the spin resonance
mode, measured in inelastic neutron scattering (INS), on the finite temperature 
self energies is investigated.    
 \end{abstract}
 \pacs{74.20.-z 74.20.Mn 74.25.Jb 74.72.-h}
  \maketitle
Recently, self consistent Eliashberg theory has been
 quite successful in fitting zero temperature
 superconductor-insulator-superconductor (SIS) point contact
 tunnel (PCT)
 junction characteristics in the high temperature superconducting cuprate Bi-2212. \cite{1}
 
 The success of that work  indicates that the
 superconducting state of the cuprates may be understood to originate from 
  a low frequency spectrum of
 spin fluctuations. This interpetation is sometimes referred to as
 the {\em pairing glue} scenario \cite{2}, and is most likely to be
appropriate for the optimal to overdoped regime of the cuprates. For an explanation of the origin of
 high temperature superconductivity in the underdoped regime, and in proximity to the insulating
antiferromagnetic phase of the cuprates, the RVB singlet picture \cite{3,31}, involving the exchange constant J in the t-J model
Hamiltonian, provides an explanation for the emergence of the high temperature superconducting phase.
 
 A strong coupling analysis has also been used  
 to interpet both angle resolved photoemission
 spectroscopy (ARPES) \cite{41,42,43}, and optical conductivity \cite{5,5a}.
 The former yields results for the single 
 quasiparticle self energy $\Sigma(\omega, T)$.
 In the latter case, 
a memory function analysis of the optical conductivity  
results in  a  frequency dependent optical self energy $\Sigma_{opt}(\omega,T)$ \cite{61}.
 
The origin of structure in the electronic spectra of the cuprates, measured in such experiments
as ARPES and optical conductivity, is a topic of much debate.   
In some analyses of ARPES \cite{6}, dispersion anomalies observed along the nodal line in
 the copper-oxide plane Brillouin Zone, have been interpeted in terms of quasiparticle scattering involving phonons, rather than
 spin fluctuations. The energy scales of
 important features in the spin fluctuation spectrum, namely the so-called spin
 resonance mode (at approximately 40 meV), and certain phonon modes in optimally doped Bi-2212
  make them potential
 candidates for understanding structure seen in ARPES. Higher energy spin
fluctuations at 80meV have been proposed for the observed dispersion anomaly in YBaCuO using an Eliashberg
analysis\cite{66}. As will be discussed later, the absence of a temperature dependence in the measured low frequency slope of the real part of $\Sigma(\omega, T)$ (Re $\Sigma(\omega, T)$) in the case of nodal ARPES measurements on Bi-2212, suggests that the spin resonance mode is not involved in nodal quasiparticle interactions in Bi-2212, similar to the situation proposed
for YBaCuO\cite{66}.

 One motivation for applying a strong coupling, Eliashberg, approach to the
 cuprates is the success of the same approach in conventional superconductors. In conventional superconductors, such as
 Pb, a self consistent Eliashberg theory, based
 on the use of a single self energy diagram, with no vertex corrections, can be
 justified by Migdal's theorem since the characteristic energy scale of phonons relative to the Fermi energy is
 very small \cite{7}. 

Another important point is that in the case of conventional superconductors, there exists a detailed, 
and quantitative description of the phonon spectral function, denoted by $\alpha^{2} F(\omega)$, 
describing the superconducting pairing mechanism. The
analog of $\alpha^{2}F(\omega)$ for the spin fluctuation spectrum in the cuprates, denoted by $I^{2} \chi(\omega)$ in
some work, is not as well understood.
The spectral function, describing the {\em pairing glue} due to spin fluctuations is modelled phenomenologically in
the present work by a prominent
 peak at the energy of the spin resonance mode 
($\Omega_{Res}$) accompanied by a broad background continuum extending up  to higher energies.\cite{7a} The value of $\Omega_{Res}$
has  been measured in inelastic neutron
 scattering (INS) experiments \cite{8}, and is about 40meV for optimally doped Bi-2212, which is 
the focus of the present work.  The ubiquitous dip feature seen in spectroscopies such as
 tunneling \cite{LC} and ARPES \cite{Dess} is taken as evidence of the key role of the spin resonance
 mode peak in the superconducting state of the cuprates.
 
 Furthermore, support for the partly phenomenological approach of the
 present work comes from extensive, ab initio, Quantum Monte Carlo numerical
 studies \cite{2,91,92,93} on the
 Hubbard and t-J models which indicate a key role is played by a low frequency spin fluctuation spectrum in
 determining the properties of the superconducting state of these models.  
 
 One of the most unusual aspects of the spin
 fluctuation physics in the cuprates is the
 decrease in intensity with increasing temperature of the spin resonance peak at $\Omega_{Res}$,
 along with its disappearence at the
 superconducting transition temperature T$_{C}$. This was first directly
 measured in INS measurements on Bi-2212.\cite{8} Similar behavior 
 has also been extracted from analysis of optical conductivity, although 
in that case the mode appears to survive, albeit significantly diminished in intensity, above T$_{C}$\cite{5a}. The emergence of such a peak in the spin fluctuation spectrum at the superconducting transition may be the result of a feedback effect into the electronic spin susceptibility due to the onset of the superconducting gap \cite{Dahm}.  
 
In the {\em pairing glue} scenario, the disappearance of the resonance mode
peak would imply that the superconducting pairing mechanism weakens with increasing 
temperature, very unlike conventional, phonon mediated, superconductors. 
The effect of a temperature dependent {\em pairing glue}
on the temperature dependence of
important quantities such as the superconducting gap, and the value of the transition
temperature T$_{C}$, is one of the subjects of the present work. It also results in significant temperature dependent 
effects in the real and imaginary parts of $\Sigma(\omega, T)$ such as
a change in slope of the Re$\Sigma(\omega, T)$ at low frequencies.
Comparison between finite temperature model calculations and
measurements of the temperature dependence of the self energy from ARPES and optical conductivity, may serve as an important  test of the validity of
the spin fluctuation {\em pairing glue} hypothesis for the cuprates.\\
\\
{\bf Theoretical Formalism}\\
\\
In the present work, a modified version of the
standard finite temperature ($T \neq 0)$ Eliashberg equations \cite{7} are solved self consistently. 
The zero temperature ($T=0$) case of these equations was used in \cite{1,LC}.
The coupled equations to be solved are given by
\begin{widetext}
\begin{eqnarray}
\Delta(\omega,T) &=& \frac{1}{Z(\omega,T)}\displaystyle\int_{0}^{\omega_{c}}d\nu 
\int_{0}^{2 \pi} \frac{d \phi}{2 \pi} c_{\Delta} {\rm Re} \bigg\{  \frac{\Delta(\nu,T) {\rm cos}^{2}(2 \phi)}{[ \nu^{2} - \Delta^{2}(\nu,T) {\rm cos}^{2}(2 \phi)]^{1/2}}\bigg\}
\displaystyle\int_{0}^{\omega_{c}}d\omega ' F(\omega ', T ) 
\nonumber \\
&\times& \bigg\{ [ n(\omega ')+f(-\nu)]
\bigg[ \frac{1}{\omega + \omega ' + \nu + i\delta} - \frac{1}{\omega - \omega ' - \nu +i\delta}\bigg] \nonumber \\
\nonumber \\
&-& [ n(\omega ') + f(\nu)] \bigg[ \frac{1}{\omega + \omega ' -\nu +i\delta} - \frac{1}{\omega - \omega ' +\nu +i\delta}\bigg ]\bigg\} 
%\nonumber\\
\end{eqnarray}
and
\begin{eqnarray}
[1-Z(\omega, T)]\omega &=& \displaystyle\int_{0}^{\infty}d\nu \int_{0}^{2 \pi} \frac{d \phi}{2 \pi} 
c_{Z} {\rm Re} \bigg \{ \frac{\nu}{[ \nu^{2} - \Delta^{2}(\nu, T){\rm cos}^{2}(2 \phi) ]^{1/2}} \bigg \}\displaystyle\int_{0}^{\omega_{max}} d\omega ' F(\omega ',T) \nonumber \\
&\times &  \bigg \{ [ n(\omega ') + f(-\nu) ] \bigg [ \frac{1}{\omega + \omega ' + \nu + i\delta}
  + \frac{1}{\omega - \omega ' -\nu + i\delta}\bigg ] \nonumber \\
 \nonumber \\
&+& [ n(\omega ') +f(\nu)] \bigg [ \frac{1}{\omega + \omega ' -\nu +i\delta}
 + \frac{1}{\omega - \omega ' + \nu + i\delta} \bigg] \bigg \} \nonumber \\
\end{eqnarray}
\end{widetext}

In the present work
\begin{equation}
\Delta(p,\omega) \; = \; \Delta(\omega,T) {\rm cos}(2 \phi)
\end{equation}
$\phi$ denotes the angular position on the Fermi surface, with the cos($2 \phi$)
dependence incorporating a d-wave symmetry superconducting state.

In place of the conventional phonon spectral function $\alpha^{2}F(\omega)$ in the Eliashberg equation
formalism, the spin
fluctuation {\em pairing glue} is described by \cite{11}
\begin{equation}
[c_{Z} \; + \; c_{\Delta} {\rm cos}(2(\phi \; - \; \phi^{'}))]F(\omega,T)
\end{equation}
The spin fluctuation spectral function $F(\omega,T)$ will be
discussed in detail in the next section.\\
\\
\\
{\bf Results}\\
\\
The results presented here are for optimally doped Bi-2212, and are an extension of our successful fitting
of the SIS tunneling conductance for this material at $T=0$.\cite{1}
The superconducting transition temperature is chosen
to be 90K in the present work. The values of $c_{Z} = 0.2 $ and $c_{\Delta} =   0.83$ were
used in reference (1) to fit the optimally doped SIS PCT experimental conductance.
The coupled Eliashberg equations for
$Z(\omega, T)$ and $\Delta(\omega, T)$ are solved self consistently 
Results for the diagonal self energy
$\Sigma(\omega,T)$ are presented and discussed later, where $\Sigma(\omega,T)$ is defined by 
\begin{equation}
\Sigma(\omega,T) \; = \; (1 \; - \; Z(\omega, T)) \omega
\end{equation}
The Re$\Sigma(\omega,T)$ is usually called the mass enhancement factor, and -2 Im$\Sigma(\omega, T)$ 
is the quasiparticle scattering rate ($1/ \tau$). 
These two quantities can be measured in APRES\cite{41,42}. Optical conductivity experiments \cite{5} yield a related optical self energy function as well. 
The temperature dependent superconducting gap, usually denoted by 
$\Delta(T)$, is defined as the frequency $\omega$ at which $\omega \; = \; {\rm Re} \Delta(\omega, T)$.

The spectral function $F(\omega, T)$ describing the {\em pairing glue} in this work is shown in Figure 1. The
 intensity of the main peak, at the resonance mode frequency $\Omega_{Res}$, decreases as the  temperature is
 increased, eventually vanishing at $T_{c}$. The  $F(\omega, T)$ curves shown in Figure 1
 incorporate the temperature dependence of the resonance mode, seen in INS and optical conductivity experiments. 
The resonance
mode rides on a weakly temperature dependent background which extends to higher frequencies.

 Figure 1 is generated using a 
model for the spectral function of the form
 \begin{equation}
 F(\omega, T)=H(T)F(\omega)^{Res}+BG(T)F(\omega)^{BG}
 \end{equation}
 The resonance mode is modelled by a Gaussian $F(\omega)^{Res}$ with peak magnitude given by
 $H(T)=H_{0}(1-T/90K)^{1/2}$, where $H_{0}$ is the initial
 peak height at T=0K.  The
temperature dependence of the peak height is similar to the BCS mean field  $\Delta(T)$ behavior, and also
 consistent with the temperature dependent mode height seen in
 INS experiments.\cite{8}
 
The background continuum is modeled using a broad 
temperature independent Gaussian function of frequency $\omega$ denoted by $F(\omega)^{BG}$ 
multiplied by a temperature dependent factor  $BG(T)=1+\alpha T/90K$, where $\alpha$ is
 adjusted to yield $\Delta(T) = 0$ at $T=90K$ in the self consistent solution of the Eliashberg equations.
 
 \begin{figure}[h]
\begin{center}
\includegraphics[scale=0.6]{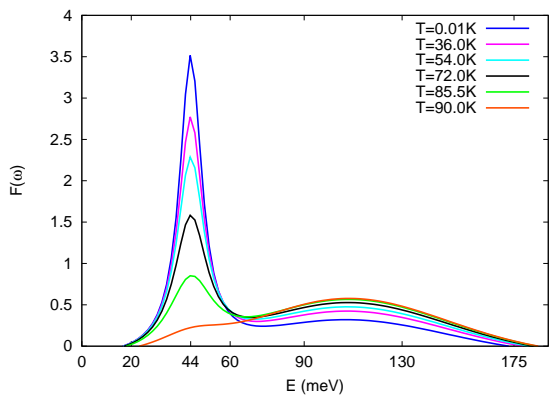}
\caption{Evolution of {\em pairing glue} spectral function $F(\omega,T)$ in equations (1) and (2) for 
various temperatures up to $T_{c}$.}
\label{a2fT}
\end{center}
\end{figure} 
 
In the present work, the resonance peak at $\Omega_{Res}$ disappears completely at $T_{C}=90$K.\cite{8} This is
consistent with INS measurements of the temperature dependence of the spin resonance mode intensity. In a recent analysis of
optical conductivity on Bi-2212 \cite{5a}, the height of the resonance mode decreases by a factor of two at the superconducting
transition temperature, with a smaller peak surviving into the normal state. The origin of this difference with
INS measurements is not understood. However, if the temperature dependence seen in optical conductivity was
used in place of the complete mode disappearance as used in the current work, the behavior discussed below
in the 2$\Delta(0)/k_{B}T_{C}$ and the self energy $\Sigma(\omega,T)$ would still be present, albeit
quantitatively different. 

In one analysis (see Fig (18) of reference (10)) of ARPES experimental data, the resonance mode
peak sits on a relatively large, temperature independent, background continuum. In that case, the decrease in the peak height, 
as the temperature is increased up to
T$_{C}$, results in a correspondingly smaller reduction in the {\em pairing glue} coupling strength compared to the current results
presented here. 

Finally, the issue of whether the area under the $F(\omega,T)$ curve should be conserved,
as the temperature changes, is unclear. This would require that area lost from the main
peak would be redistributed to higher frequencies. However, even if this requirement were imposed,
there would still likely be a reduction in the strength of the {\em pairing glue} (see equation (7) ahead) with increasing temperature.  

The consequences of a complete disappearance of the resonance mode peak are investigated in
this work, but uncertainties 
in the frequency dependence of the {\em pairing glue} spectral function should be kept in mind.

One of the interesting features of Bi-2212 (optimally doped) is the high value of the ratio
$2 \Delta(0)/ k_{B} T_{C}$, which is observed to be 9.3. Table 1
lists values from reference (23) of measured $T_{C}$ and zero temperature gap values ($\Delta(0)$) and the corresponding $2 \Delta(0)/ k_{B} T_{C}$.

\begin{table}[h]
\caption{$T_{C}$ and $\Delta(0)$ values from reference (23) }
\centering                          % used for centering table
\begin{tabular}{c c c}  
& & \\          % centered columns (3 columns)
\hline\hline   
 & & \\                     % inserts double horizontal lines
$T_{C}$(K)\ & $\Delta(0)$ (meV) \ & $2 \Delta(0)/ k_{B} T_{C}$ \\ [0.5ex] % inserts table heading
& & \\
\hline     
& & \\                         % inserts single horizontal line below heading
51 & 10.5 & 4.78    \\              % inserting the body of the table
62 & 17.5 & 6.56 \\
92 & 31  &   7.83  \\
95 & 38   &   9.3 \\ [1ex]         % [1ex] adds vertical space
\hline                              % inserts single line
\end{tabular}
\label{ratio}                % is used to refer this table in the text
\end{table}
 
 The disappearance of the resonance mode at the measured $T_{C}$ may
 be an important contribution to the large $2\Delta(0)/k_{B}T_{c}$ ratios in Table 1.

To illustrate the effect of assuming a temperature independent {\em pairing glue},
Figure 2 shows the resulting $\Delta(T)$ from using the $T = 0$K $F(\omega, T)$ from Figure 1 for all finite
temperatures in the Eliashberg equations. 
The calculated $T_{c}$ for this case is 153K as seen in Figure 2, much higher
 than the measured 90K value for optimally doped Bi-2212. 
 The value of the ratio $2\Delta(0)/k_{B}T_{c}$ is 5.7,
 much less than the observed value of 9.3. This same discrepancy
will occur in the case of the other Bi-2212 doping levels previously analysed\cite{1}.
 This  observation implies that a temperature dependent weakening of
the  strength of the {\em pairing glue}, and a consequent reduction in the
transition temperature $T_{C}$,  may be at the origin of the large ratio values shown in Table 1
Furthermore, the trend seen in Table 1 for the  $2\Delta(0)/k_{B}T_{c}$ ratio to decrease
with overdoping is also consistent with the decreasing prominence of the resonance mode 
in the {\em pairing glue} spectrum with overdoping \cite{1}.
 \begin{figure}[htb]
\begin{center}
 \includegraphics[scale=0.6]{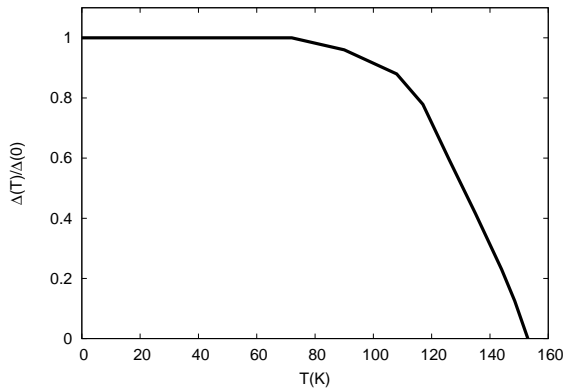}
 \caption{$\Delta(T)/\Delta(0)$ vs $T$ using the $F(\omega,T)$ for $T=0$ from Figure (1).
 $T_{c}$ is 153K.}
 \label{modetempind}
 \end{center}
 \end{figure}

Figures (3) and (4) show the
Re$\Sigma(\omega,T)$ and Im$\Sigma(\omega,T)$ for the case where 
$F(\omega,T)$ is kept at its $T \; = \; 0$ shape for all temperatures up
to the transition temperature of 153K.
As temperature increases, the peak structures in the self
 energy functions are smeared out due to the Fermi and Bose
 functions in the Eliashberg equations. It is worth noting that, in the d-wave superconducting state, the main peak in
the Re$\Sigma(\omega,T=0)$ is located at approximately 75 meV, 
slightly below the value $\Delta(0) \; + \; \Omega_{Res} \; = \; 82$ meV (using $\Delta(0)=38$meV
and $\Omega_{Res}=44$ meV \cite{1}),
which would be expected for an s-wave superconductor. 

A measure of the {\em pairing glue} coupling constant involved in determining $\Sigma(\omega,T)$, which is denoted by $\lambda$ in
the present work, can be extracted in two ways. One is by analogy with the electron-phonon
coupling constant in conventional superconductors, by use of
\begin{equation}
\lambda \; = \; 2 c_{Z} \int_{0}^{\infty} F(\omega,T)/ \omega
\end{equation}
A second measure of $\lambda$ is from the slope, at low $\omega$, of Re$\Sigma(\omega,T)$. 
Results for $\lambda$ from both of these procedures are tabulated in Table II for the
case of a temperature dependent {\em pairing glue}, which will be discussed later.
For the current situation where the $F(\omega,T)$ is assumed to stay constant at its $T=0$K
shape, 
the slope of Re$\Sigma(\omega,T)$
in Figure (3) is constant at low $\omega$, and consequently $\lambda$ does not change with temperature.
 
 \begin{figure}[htb]
\begin{center}
 \includegraphics[scale=0.6]{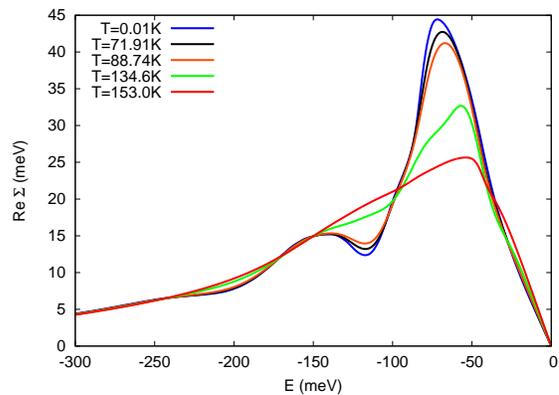}
 \caption{Temperature dependence of the real part of the self energy
 $\Sigma(\omega)$ using the $T=0$ $F(\omega,T)$ from Figure (1) for all temperatures
shown in this figure.}
 \label{resigtempind}
 \end{center}
 \end{figure}

 \begin{figure}[htb]
\begin{center}
\includegraphics[scale=0.6]{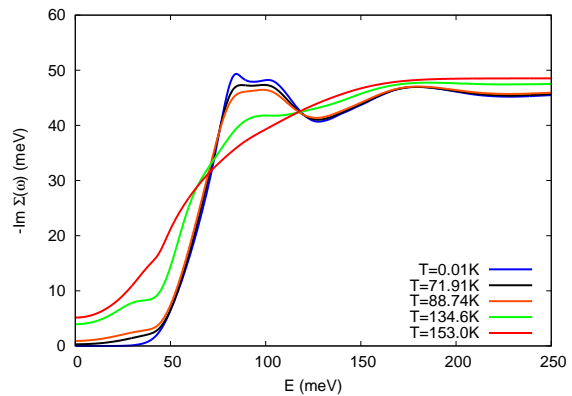}
\caption{Temperature dependence of $-{\rm Im} \Sigma(\omega)$ for same conditions as in Figure (3).}
\label{imsigtempind}
\end{center}
\end{figure}

It is interesting to compare the temperature evolution seen in  Figures (3) and (4) with that of Pb for which the 
Re$\Sigma(\omega, T)$ and Im$\Sigma(\omega, T)$ are shown in Figures (5) and (6). The lead gap 
$\Delta(0) = 1.35$meV
is almost a factor of 3 less than the first phonon peak of approximately 4.4meV in the lead phonon $\alpha^{2}F(\omega)$ spectral
weight, which itself is temperature independent. The result is a relatively undramatic temperature
evolution in the Pb Re$\Sigma(\omega,T)$ and Im$\Sigma(\omega,T)$ compared with that seen in
Figures (3) and (4) where the value of $\Delta(0)$ and $\Omega_{Res}$ are comparable. 

 \begin{figure}[h]
\begin{center}
\includegraphics[scale=0.6]{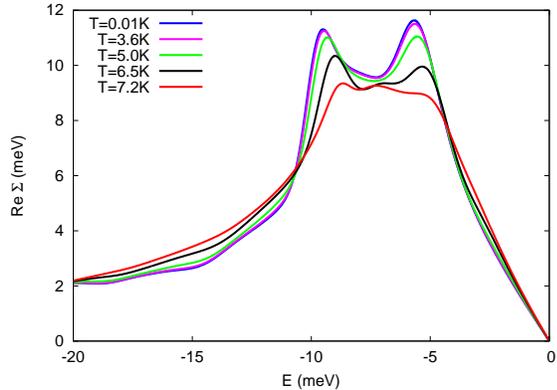}
\caption{Temperature evolution of Re$\Sigma(\omega,T)$ for lead (Pb)}
\end{center}
\end{figure}

\begin{figure}[h]
\begin{center}
\includegraphics[scale=0.6]{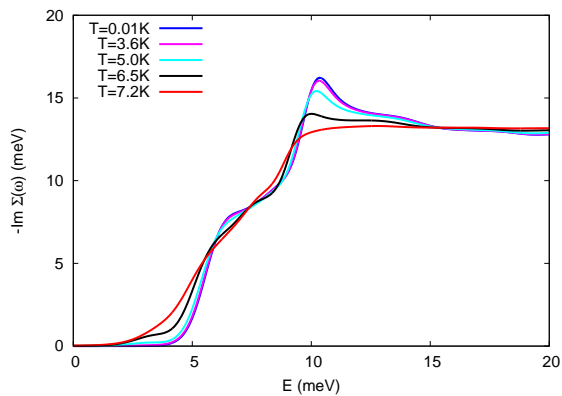}
\caption{Temperature evolution of -Im$\Sigma(\omega,T)$ for lead (Pb)}
\end{center}
\end{figure}
 
Figures (7), (8) and (9) show results for $\Delta(T)$, Re$\Sigma(\omega,T)$ and Im$\Sigma(\omega,T)$ 
for the case of the temperature dependent $F(\omega,T)$ shown in Figure (1). The superconducting gap $\Delta(T)$ now goes to zero at 90K,
due to the choice of $\alpha=0.6$ in the background continuum $BG(T)$ function. 
The $T=0$ value for $\Delta(0)=38$meV
is in agreement with measured value from SIS PCT measurements \cite{12}, and so the correct $2 \Delta(0)/k_{B}T_{C}$ 
of 9.3 is reproduced. The overall temperature dependence of $\Delta(T)$ in Figure (4)  resembles that of the resonance mode peak height temperature dependence 
used for $H(T)$.

\begin{figure}[h]
\begin{center}
\includegraphics[scale=0.6]{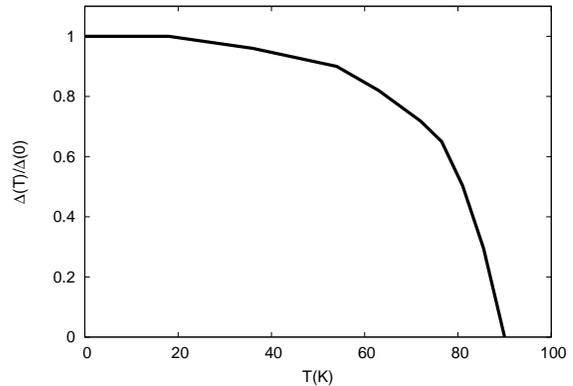}
\caption{Temperature dependent superconducting gap $\Delta(T)$ for optimal doped Bi-2212 for
the temperature dependent {\em pairing glue} spectra $F(\omega,T)$ shown in Figure (1)}
\label{gapTbi2212}
\end{center}
\end{figure}

\begin{table}[h]
\caption{Coupling strength of optimally doped Bi2212 for various temperatures up to $T_{c}$ (from Figure (8)) $c_{Z} = 0.2$ (see text)}
\centering                          % used for centering table
\begin{tabular}{c c c}  
& & \\          % centered columns (3 columns)
\hline\hline   
 & & \\                     % inserts double horizontal lines
T(K)\ & $\frac{\partial \Sigma(\omega)}{\partial \omega}$\ & $\lambda$ \\ [0.5ex] % inserts table heading
& & \\
\hline
& & \\                              % inserts single horizontal line below heading
0.01 & 0.46 & 0.59    \\              % inserting the body of the table
36 & 0.45 & 0.57 \\
54 & 0.44   &   0.54  \\
72 & 0.41   &   0.49 \\
85.5 & 0.36  &   0.41 \\
90 & 0.23    & 0.29  \\ [1ex]         % [1ex] adds vertical space
\hline                              % inserts single line
\end{tabular}
\label{couplingstemp}                % is used to refer this table in the text
\end{table}

Figures (8) and (9) show the corresponding temperature dependent real and
imaginary parts of $\Sigma(\omega,T)$. There are significant differences between these results
and those shown in Figures (3) and (4). The Re$\Sigma(\omega, T)$ of Figure (8) show a much more
rapid decrease in the main peak at approximately $\Omega_{Res}+\Delta(T)$ which reflects the
decrease in the intensity of the resonance mode with increasing temperature. 
The dip feature also fills in more rapidly
compared to the temperature evolution of Figure (3). Furthermore, there is now a noticeable
decrease in the slope of Re$\Sigma(\omega,T)$ at low frequencies, reflecting the decrease in the
effective coupling constant ($\lambda$) as the resonance mode disappears. This is illustrated in the accompanying
Table II. 

 In Figure (9), the disappearance of the spin resonance mode peak with increasing temperature results
 in a significant reduction in Im$\Sigma(\omega,T)$ in the energy range 50 meV to 100meV. This energy
range brackets the energy $\Omega_{Res}+\Delta(T)$ where the mode shows up in such quantities as the
self energy in Eliashberg theory. A significant reduction in the measured width of the MDC peak in 
the same frequency range, measured by ARPES,
could search for this effect.
 
A limited amount of experimental measurements of the real and imaginary single particle $\Sigma(\omega,T)$,
have been  extracted from
 ARPES MDC curves on Bi-2212 \cite{41,42}. Optical conductivity experiments on Bi-2212 \cite{5} have
also yielded an optical self energy $\Sigma_{opt}(\omega,T)$, 
along with a temperature dependent spectral
function for the underlying modes interacting with the quasiparticles. 
These analyses appear to reveal several different contributions
to the extracted self energies.  Separating out different contributions to the real and imaginary components of the
quasiparticle self energy has been the subject of recent work (see Figure (10) of reference (10)) and reference (24).

\begin{figure}[h]
\begin{center}
\includegraphics[scale=0.6]{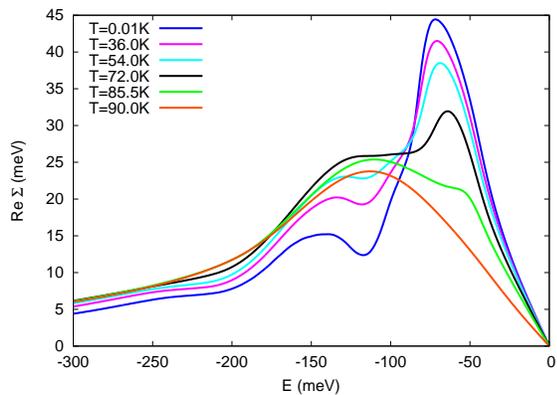}
\caption{Temperature dependence of Re$\Sigma(\omega)$ for the temperature dependent $F(\omega,T)$
spectra shown in Figure (1).}
\label{ReSigTemp}
\end{center}
\end{figure}

\begin{figure}[h]
\begin{center}
\includegraphics[scale=0.6]{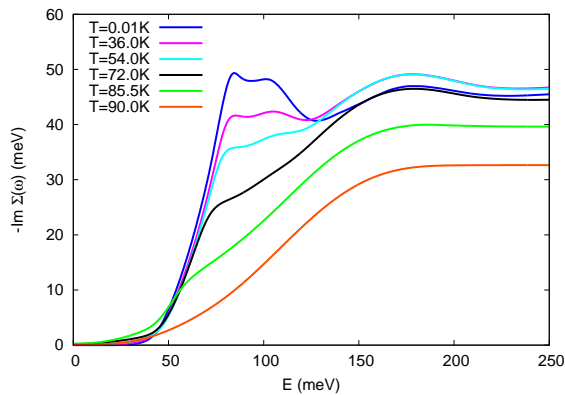}
\caption{Im$\Sigma (\omega)$ for same conditions as in Figure (8).}
\label{imsig3}
\end{center}
\end{figure}

In reference(8), a contribution is
present in $1/ \tau$ = -2Im$\Sigma_{opt}(\omega,T)$ which increases linearly with energy, and
is temperature dependent. The linear behavior with energy in $1 / \tau$ 
has been interpeted as evidence for the Marginal Fermi Liquid hypothesis
\cite{Varma}. Furthermore,
a large energy and temperature dependent background is measured in the Re$\Sigma(\omega,T)$ in reference (6) which is
subtracted off to yield the temperature dependent contribution to Re$\Sigma(\omega,T)$.
These contributions complicate interpetation of the ARPES and optical conductivity data, and their comparison with the 
present model. In the analyses of ARPES and optical conductivity experiments, a background
continuum extending up to 400meV is present in the $I^{2} \chi(\omega)$ curves (see Figures (18) and (31)
of reference (10)), in contrast to the {\em pairing glue} spectral curves shown in Figure (1) of this
work where the high energy continuum extends up to 175 meV approximately.
 The linear increase in $1/ \tau$ with increasing energy in reference (8) most likely results from this. 
  
Bearing these issues in mind, figures 3(b) and 4(b) of reference (6), which show
the MDC linewidth and the Re$\Sigma(\omega,T)$ for nodal line ARPES of Bi-2212 respectively,
for different temperatures, do not appear to display the rapid decrease in linewidth or
a discernible change in slope for Re$\Sigma(\omega,T)$ at low frequencies, that are shown in
Figures (9) and (8) of the present model calculations. This could be interpeted as evidence that
the nodal Bi-2212 quasiparticles are not coupling to the 40meV temperature dependent spin
resonance mode. The zero superconducting gap for these quasiparticles would point to a central role for the 40meV mode in the superconducting pairing in Bi-2212. Experimental studies of
the anti-nodal self-energies would be of interest in following up on this issue.
   
In conclusion, results have been presented for an finite temperature Eliashberg calculation of a d-wave
superconducting case with  temperature dependent {\em pairing glue}. The calculated real and imaginary diagonal
self energies may prove a useful probe of the {\em pairing glue} description of high temperature
superconductors when compared with ARPES and optical conductivity experiments.

The authors wish to acknowledge useful conversations with J. F. Zasadzinski

\end{document}